\documentclass[preprint,5p,times,
sort&compress,
twocolumn
]{elsarticle}
\usepackage[utf8]{inputenc}

\usepackage{amsthm, amsfonts, amssymb, comment}	
\usepackage[fleqn]{amsmath}				
\usepackage{graphicx}					
\usepackage{txfonts}					
\usepackage{hyperref,xcolor}			
\usepackage{geometry}                   
\usepackage{lineno}						
\modulolinenumbers[1]					
\usepackage{mathtools, cuted}           
\usepackage{xcolor}
\usepackage[normalem]{ulem}

\newcommand{\newc}{\newcommand}
\newcommand{\ben}{\begin{eqnarray}}
\newcommand{\een}{\end{eqnarray}}
\newc{\ra}{\rightarrow}
\newc{\bfx}{{\bf x}}
\newc{\bfV}{{\bf V}}
\newc{\cO}{{\cal O}}
\newc{\bfv}{{\bf v}}
\newc{\bfu}{{\bf u}}
\newc{\bfp}{{\bf p}}
\newc{\ve}{{\varepsilon}}
\newc{\Psibar}{\overline\Psi}
\newc{\w}{{\bf w}}
\newc{\E}{{\mathbf{E}}}
\newc{\EE}{{\mathcal E}}
\newc{\bfn}{{\mathbf\nabla}}
\newc{\la}{{\cal L}}
\newc{\tla}{{\tilde{\cal L}}}
\newc{\bp}{{\bf p}}
\newc{\ho}{\hookrightarrow }
\newc{\bP}{{\bf P}}
\newc{\pd}{{\partial}}
\newc{\piv}{{\partial_4}}
\newc{\pv}{{\partial_5}}
\newc{\bJ}{{\bf J}}
\newc{\bze}{{\mathbf 0}}
\newc{\bK}{{\bf K}}
\newc{\tphi}{{\tilde\phi}}
\newc{\tF}{{\tilde F}}
\newc{\tD}{{\tilde D}}
\newc{\tJ}{{\tilde J}}
\newc{\tj}{{\tilde j}}
\newc{\bD}{{\bf D}}
\newc{\tvphi}{{\tilde\varphi}}
\newc{\trho}{{\tilde\rho}}
\newc{\ttheta}{{\tilde\theta}}
\newc{\tpsi}{{\tilde\psi}}
\newc{\tu}{{\tilde u}}
\newc{\cD}{{\cal D}}
\newc{\tPhi}{{\tilde\Phi}}
\newc{\tPsi}{{\tilde\Psi}}
\newc{\tA}{{\tilde A}}
\newc{\talpha}{{\tilde\alpha}}
\newc{\tbeta}{{\tilde\beta}}
\newc{\bA}{{\mathbf A}}
\newc{\bB}{{\bf B}}
\newc{\br}{{\bf r}}
\newc{\sig}{{\mathbf\sigma}}
\newc{\eg}{{\rm e.g.\ }}
\newc{\ie}{{\rm i.e.\ }}
\newcommand{\bey}{\begin{eqnarray}}
\newcommand{\pslash}{\not{\hbox{\kern-2.3pt $p$}}}
\newcommand{\pdslash}{\not{\hbox{\kern-2pt $\partial$}}}
\newcommand{\eey}{\end{eqnarray}}

\begin{document}

		\title{Partition function for position-dependent mass systems from 
        superestatistics 
}
		
	\author[itapetinga,profisica]{Ignacio S. Gomez\corref{cor1}}
\ead{ignacio.gomez@uesb.edu.br}

\cortext[cor1]{Corresponding author}
\address[itapetinga]{Departamento de Ciências Exatas e Naturais, Universidade Estadual do Sudoeste da Bahia,
			    BR 415, Itapetinga - BA, 45700-000, Brazil}
\address[profisica]{PROFÍSICA – Programa de Pós-Graduação em Física, 
Universidade Estadual de Santa Cruz, 45650-000 Ilhéus, BA, Brazil}

\author[itapetinga]{Matheus Gabriel Alves Santos}
\ead{202310403@uesb.edu.br}

\author[itapetinga]{Daniela de Almeida dos Santos}
\ead{202010666@uesb.edu.br}

\author[itapetinga,profisica]{Ronaldo Thibes}
\ead{thibes@uesb.edu.br}

\begin{abstract}
In this work we show a connection between  
superestatistics and position-dependent mass (PDM) systems in the context of the canonical ensemble.
The key point is to set the fluctuation distribution of the inverse temperature in terms of the PDM of the system. 
For PDMs associated to Tsallis and Kaniadakis nonextensive statistics the pressure and entropy of ideal gas result lower than the standard case but maintaining monotonic behavior. 
Gas of non-interacting harmonic oscillators provided with quadratic and exponential PDMs exhibit a behavior of standard 3D harmonic oscillator gas and a linear specific heat respectively, the latter being consistent with Nernst's third law of thermodynamics.
Thus, a combined PDM-superestatistics scenario 
offers an alternative way to study the effects of the inhomogeneities of PDM systems in their thermodynamics.
\end{abstract}
		
\begin{keyword}
superestatistics \sep position-dependent mass \sep canonical ensemble \sep partition function \sep nonextensive statistics \sep deformed algebraic structures

\end{keyword}

\maketitle

\section{Introduction}

Superestatistics is a branch of statistical mechanics that deals with nonequilibrium and nonlinear systems\cite{SUPER}. The main assumption of this approach is the existence of a long-term stationary state having a intensive quantity fluctuating spatially and temporally, in such a way that averaging over the fluctuations it is possible to obtain a wide set of general statistics, including nonextensive ones as the Tsallis statistics. In addition, high-order correction terms and different behaviors from large variance can be obtained by application of this superposition of statistics, briefly called \emph{superestatistics}. 
Examples of phenomena obeying more general statistics than the Tsallis one that are well described by superestatistics can be found in the velocity fluctuations in a Taylor-Couette flow\cite{Jung}, fusion plasma physics\cite{Sattin} and Lagrangian turbulence\cite{Lagrangian}, among others. Also, nonthermal and suprathermal distributions\cite{Ourabah}, stretched exponentials\cite{stretched}, and generalized entropies\cite{Hanel} have been studied in the context of superestatistics.  

On the other hand, position-dependent mass (PDM) systems have arisen originally to describe transport phenomena in semiconductor heterostructures \cite{PDM1,PDM2} with subsequent applications in DFT \cite{PDM-DFT}, SUSY quantum mechanics \cite{PDM-SUSY}, nuclear physics \cite{PDM-NP}, nonlinear optics \cite{PDM-NO}, Landau quantization \cite{PDM-LQ}, superestatistical partition functions \cite{Maike-2021} and group entropy algebraic structures linked with PDM Schrödinger equations \cite{Gomez-2021}, among other fields. Recently, intimately links between canonical transformations, generalized derivative operators and nonextensive algebraic structures connected with PDM systems have been addressed \cite{Gomez-2021,Morse-EPL,daCosta-PRE-2020}.

In this work, by means of a suitable choice for the fluctuation distribution of the inverse temperature, we characterize the canonical ensemble partition function of one-dimensional PDM systems in terms of a superestatistical partition function. Thus, we show that superestatistics can describe the thermodynamics of PDM systems in a simple and unified way. The work is organized as follows. In Section 2 we give the minimal preliminaries for the understanding of our proposal. 
In Section 3 we establish the theoretical connection between PDM canonical ensemble partition functions and superestatistical partition functions. 
Next, in Section 4, we illustrate the formalism presented with some examples. 
Finally, in Section 5 some conclusions and perspectives are outlined.
\section{Preliminaries}

We start with a review of the minimal concepts and definitions required for the development of the subsequent sections of this work. 

\subsection{
Position-dependent mass systems}

The progress of nonextensive statistics \cite{Tsallis-1988, Tsallis-Bukmann-1996, Baldovin-Robledo-2002, Kaniadakis-2002, Kaniadakis-2005, Tsallis-2009, Tirnakli-Borges-2016} has driven the development of some of their associated mathematical structures, in particular
the $\kappa$-algebra \cite{Kaniadakis-2002,Kaniadakis-2005} and
the $q$-algebra, \cite{Nivanen-Mehaute-Wang-2003,Borges-2004} respectively related to Kaniadakis and Tsallis statistics.
The prescriptions for the position-dependent masses (PDM) $m_{q,\kappa}(u)$,
associated to the $q$-algebra and $\kappa$-algebra, allow us to define arbitrary deformations $\eta=\eta(x)$ 
of the position $x$ given by
\cite{daCosta-PRE-2020}
\begin{equation}\label{PDM-prescription}
    \textrm{d}\eta=\sqrt{\frac{m(x)}{m_0}}\textrm{d}x.
\end{equation}
For the Tsallis PDM $m_q(x)$ case, \eqref{PDM-prescription} 
establishes a canonical trasformation map between the quantum harmonic oscillator 
with deformed momentum operator and the Morse quantum oscillator \cite{Morse-EPL}.
Replacing $\eta(x)$ by the $q,\kappa$-differentials of the $q,\kappa$-algebra \cite{Borges-2004,Kaniadakis-2002} 
    $d_q x = \frac{dx}{1+(1-q)x}$ 
     and   $d_\kappa x = \frac{dx}{\sqrt{1+\kappa^2x^2}}$
in \eqref{PDM-prescription},
we obtain
\begin{subequations}
\label{PDMs}
\begin{align}
    m_{q}(x)&=\frac{m_0}{\left[1+(1-q)x \right]^2},\\
    m_{\kappa}(x) &=\frac{m_0}{1+\kappa^2 x^2},
\end{align}
\end{subequations}
for the PDMs of the $q,\kappa$-algebras 
with $m_0$ denoting the constant mass case. 
From the $q,\kappa$ differentials 
follow the $x_{q,\kappa}$ deformed numbers \cite{Borges-2004,Kaniadakis-2002}
\begin{subequations}\label{deformation-number}
\begin{align}
        x_q &= \int dx_q = \frac{1}{1-q}\ln(1+(1-q)x)=\ln (e_q(x)),\\
x_\kappa &= \int dx_\kappa = \frac{1}{\kappa}  \operatorname{arcsinh} {(\kappa x)} = \ln (e_\kappa(x)),
\end{align}
\end{subequations}
with $e_{q,\kappa}(x)$ denoting the $q,\kappa$ deformed exponential functions associated to the $q,\kappa$-algebra. 
It is worth mentioning that, by taking the limits $q\rightarrow1$ and $\kappa\rightarrow0$ in the above definitions, the $q,\kappa$-differentials, the $q,\kappa$ deformed numbers, the $q,\kappa$ exponentials and the PDMs recover the usuals differential $dx$, number $x$ and exponential $e^x$ as well as the constant mass $m_0$, thus emphasizing the alternative role of the entropic indexes $q,\kappa$ as deformation parameters.

\subsection{Superestatistical partition function and thermodynamics}
For the sake of simplicity, hereafter, we restrict our analysis to classical systems provided with a continuous phase space $\Gamma$ whose points are denoted by $(x,p)$. 
The superestatistical partition function of a given system is given by \cite{SUPER}
\begin{equation}\label{SPF}
    \mathcal{Z}= \int B(E)dE ,
\end{equation}
with $B(E)$ the superestatistical Boltzmann factor 
\begin{equation}\label{SBF}
    B(E)=\int_0^{\infty} f(\beta)e^{-\beta E}d\beta.
\end{equation}
Here, $f(\beta)$ is the distribution of the local inverse temperature $\beta$, containing information about the macroscopic fluctuations of $\beta$. As a special case, if we substitute $f(\beta)=\delta (\beta - \beta_0)$ in \eqref{SBF} with $\beta_0$ a constant value, we get 
\begin{equation}
    Z= \int e^{-\beta_0 E}dE,
\end{equation}
which is nothing but the canonical ensemble partition function of a system at equilibrium with a reservoir of constant temperature 
$T=(k_B\beta_0)^{-1}$, with $k_B$ denoting the Boltzmann constant. The Maxwell relations allow us to obtain the thermodynamic functions
\begin{subequations}
    \begin{align}
        A=-\frac{1}{\beta_0}\ln Z \,,\\
        P = -\Bigg(\frac{\partial A}{\partial V}
        \Bigg)_T \,, \\
        S=-\Bigg(\frac{\partial A}{\partial T}
        \Bigg)_V \,,\\
        G=A+PV\,, \\
        U = A+TS\,, \\
        C=\Bigg(\frac{\partial U}{\partial T}
        \Bigg)_V\,,
    \end{align}
\end{subequations}
for the Heltmholtz free energy, pressure, entropy, Gibbs free energy, internal energy and heat capacity, respectively. 

\section{PDM canonical partition function from superestatistical partition function}
Next, we give the connection between the superstatistical partition function \eqref{SPF} and the canonical partition function of a PDM system. 
For simplicity, we consider one-dimensional systems provided with a position-dependent mass $m(x)$. 
Let such a system be composed of PDM particles with an arbitrary PDM $m(x)$ subjected to a potential $V(x)$ (i.e, each particle experiments a force $F=-dV/dx$) and at equilibrium with a reservoir of constant temperature $T=(k_B\beta_0)^{-1}$.
Assuming that the system is composed by $N$ non-interacting particles, its canonical ensemble partition function is given by 
\begin{eqnarray}\label{PDM-PF}
    Z= \frac{1}{h^{N} N!}Z_1^N\,,
\end{eqnarray}
with 
\begin{equation}\label{single-PDM-PF}
    Z_1=
    \int\int e^{-\beta_0 \Big(\frac{p^2}{2m(x)}+V(x)\Big)}dxdp=
    \sqrt{\frac{2\pi}{\beta_0}}
    \int \sqrt{m(x)}e^{-\beta_0 V(x)}dx
\end{equation}
the single PDM partition function and $h$ denoting Planck's constant.

Now we consider a system composed by $N$ non-interacting particles of constant mass $m_0$ under the influence of a potential $\widetilde{V}(x)=\frac{m(x)}{m_0}V(x)$. Then, by choosing 
\begin{eqnarray}\label{choice}
    f(\beta)=\delta(\beta - \gamma(x)) \quad \textrm{with} \quad  \gamma(x)=\beta_0\frac{m_0}{m(x)}, 
\end{eqnarray}
the superestatistical partition function results to $\mathcal{Z}=(1/(h^N N!))\mathcal{Z}_1^N$ with $\mathcal{Z}_1$ expressed by 
\begin{eqnarray}\label{connection}
    \mathcal{Z}_1=\int B(E)dE=
    \int dE\int_0^{\infty} d\beta \delta(\beta - \gamma(x))e^{-\beta E}
    =\nonumber\\
    \int\int 
    e^{-\beta_0\frac{m_0}{m(x)}\Big(\frac{p^2}{2m_0}+\widetilde{V}(x)\Big)}
    dxdp=
    \int\int e^{-\beta_0 \Big(\frac{p^2}{2m(x)}+V(x)\Big)}dxdp =Z_1 \,,
\end{eqnarray}
which is precisely the single PDM partition function \eqref{single-PDM-PF}.
Equation \eqref{connection} demonstrates that for non-interacting particles, we can reproduce the canonical partition function of a position-dependent mass particle from the superestatistical partition function \eqref{SPF}, the effective potential $\widetilde{V}(x)=\frac{m(x)}{m_0}V(x)$ and 
the choice \eqref{choice}. Moreover, we have the biunivocal correspondence between pairs of masses and potentials in the PDM and superestatistical frames:
\begin{eqnarray}\label{pairs}
    &(m(x),V(x)) \Longleftrightarrow (m_0,\widetilde{V}(x))& \nonumber\\ 
    &\textrm{with} \quad   f(\beta)=\delta\Big(\beta - \beta_0\frac{m_0}{m(x)}\Big) 
    \quad , \quad
    \widetilde{V}(x)=\frac{m(x)}{m_0}V(x)&.
\end{eqnarray}

\section{Applications: thermodynamics from PDM partition functions}

In the following, we study some particular cases that result from combining $m(x)$ and $V(x)$. Since we are considering $N$ non-interacting particles, it is sufficient to calculate the single partition function $Z_1$ and then use the factorization for the total partition function $Z=(1/(h^NN!))Z_1^N$. 

\subsection{Case 1: PDM ideal gas}

By substituting $V(x)=0$ in \eqref{single-PDM-PF}, we obtain
\begin{eqnarray}\label{case-free-single}
   Z_1= \sqrt{\frac{2\pi}{\beta_0}}
    \int_0^{L} \sqrt{m(x)}dx =
    \sqrt{\frac{2\pi m_0}{\beta_0}} \int_0^{L} d\eta
    =
    \sqrt{2\pi m_0k_BT} L_\eta\,,
\end{eqnarray}
where we have used the fact that $L_\eta=\eta(L)$ is the transformation \eqref{PDM-prescription}
of the length $L$ of the gas for an arbitrary PDM $m(x)$. Furthermore, in order to avoid divergences in the integral $\int \sqrt{m(x)}$, we assume that the particles are confined to a finite volume $x\in [0,L]$.
Identifying $V=L$ (so $V_\eta=\eta(V)$) as the volume of the gas, from \eqref{PDM-PF} and \eqref{case-free-single}  setting $\sqrt{2\pi m_0k_B T_0}V_0/h=1$ for  a reference temperature $T_0$ and a characteristic macroscopic volume $V_0$, the PDM partition function of the system reads
\begin{eqnarray}\label{PDM-ideal-gas}
    Z=\frac{1}{N!}\Bigg(\sqrt{\frac{T}{T_0}}\frac{V_\eta}{V_0}\Bigg)^N\,.
\end{eqnarray}
It is clear from \eqref{PDM-ideal-gas} that we can recover the classical one-dimensional ideal gas in the limit
of constant mass $m(x)\rightarrow m_0$ ($V_\eta\rightarrow V$)
\begin{eqnarray}\label{ideal-gas}
    Z_{\textrm{ideal gas}}=\frac{1}{N!}\Bigg(\sqrt{\frac{T}{T_0}}\frac{V}{V_0}\Bigg)^N.
\end{eqnarray}

\subsubsection{Tsallis PDM ideal gas}
From the $q$-number definition  (3a) and \eqref{PDM-ideal-gas}, it is immediate to obtain 
\begin{eqnarray}\label{free-PF-q}
Z_{q} 
=
\frac{1}{N!}\Bigg(\sqrt{\frac{T}{T_0}}\Bigg(\frac{\ln (1+(1-q)\frac{V}{V_0})}{1-q}\Bigg)\Bigg)^N,
\end{eqnarray}
for the PDM ideal gas Tsallis partition function. 
Then, the thermodynamic functions (7a)--(7f) result in
\begin{subequations}
    \begin{align}
        A_q(V,T)=-k_BT \log \left(\frac{\left(\frac{\sqrt{T/T_0} \log (1+(1-q)V/V_0)}{1-q}\right)^N}{N!}\right) \,, \\
        P_q(V,T)=\frac{Nk_BT(1-q)}{V_0(1+(1-q)V/V_0) \log (1+(1-q)V/V_0)} \,,\\
        S_q(V,T)=k_B\log \left(\frac{\left(\frac{\sqrt{T/T_0} \log (1+(1-q)V/V_0)}{1-q}\right)^N}{N!}\right)+k_B\frac{N}{2}\,,\\
    G_q(V,T)=A_q(V,T)+P_q(V,T)V\,,
        \\
        U_q(V,T)=\frac{Nk_BT}{2}\,,\\
        C_q(V,T)=\frac{Nk_B}{2}\,.
    \end{align}
\end{subequations}

\subsubsection{Kaniadakis PDM ideal gas}
Similarly, from the $\kappa$-number definition  (3b) and \eqref{PDM-ideal-gas}, it is immediate to obtain 
\begin{eqnarray}\label{free-PF-q}
Z_{\kappa} =
\frac{1}{N!}\Bigg(\sqrt{\frac{T}{T_0}}\Bigg(\frac{\textrm{arcsinh}\Big(\kappa \frac
{V}{V_0}\Big)}{\kappa}\Bigg)\Bigg)^N,
\end{eqnarray}
for the PDM ideal gas Kaniadakis partition function. 
In that case, the thermodynamic functions (7a)--(7f) result in
\begin{subequations}
    \begin{align}
        A_\kappa(V,T)= -k_BT \log \left(\frac{\left(\frac{\sqrt{T/T_0} \textrm{arcsinh}(\kappa  V/V_0)}{\kappa }\right)^N}{N!}\right)\,,\\
        P_\kappa(V,T)=
        \frac{\kappa  N k_BT}{V_0(\sqrt{\kappa ^2 (V/V_0)^2+1)} \textrm{arcsinh}(\kappa  V/V_0)}\,,
        \\
        S_\kappa(V,T)=
        k_B\log \left(\frac{\left(\frac{\sqrt{T/T_0} \textrm{arcsinh}(\kappa  V/V_0)}{\kappa }\right)^N}{N!}\right)+k_B\frac{N}{2}\,,
        \\
        G_\kappa(V,T)=A_\kappa(V,T)+P_\kappa(V,T)V\,,
        \\U_\kappa(V,T)=\frac{Nk_BT}{2}\,,\\        C_\kappa(V,T)=\frac{Nk_B}{2}\,.
    \end{align}
\end{subequations}
In Fig 1, we show the isothermals and the entropy per particle for the Tsallis and Kaniadakis PDM ideal gas. In both cases, the effect of the PDM is to diminish the pressure in the expansion regarding the constant mass case. In turn, the PDM entropy turns out to be lower than in the constant mass case. Differences between Tsallis and Kaniadakis cases are only observed for the pressure in the interval $V/V_0\in [0,4]$.  The PDM Tsallis and Kaniadkais entropies result superposed in the studied range.  
\begin{figure}[th]
\centerline{\includegraphics[width=9cm]{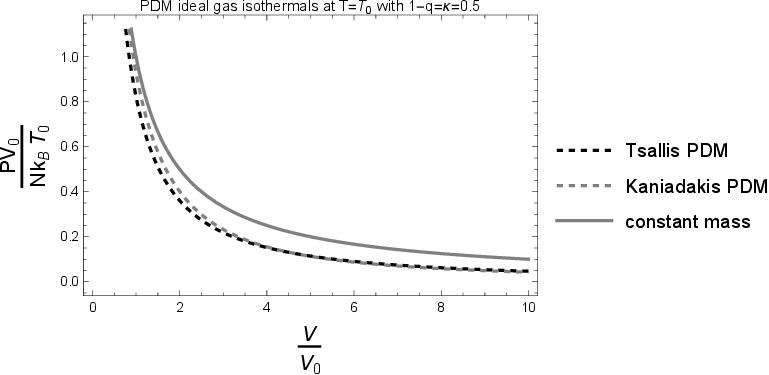}}
\vspace*{8pt}
\centerline{\includegraphics[width=9cm]{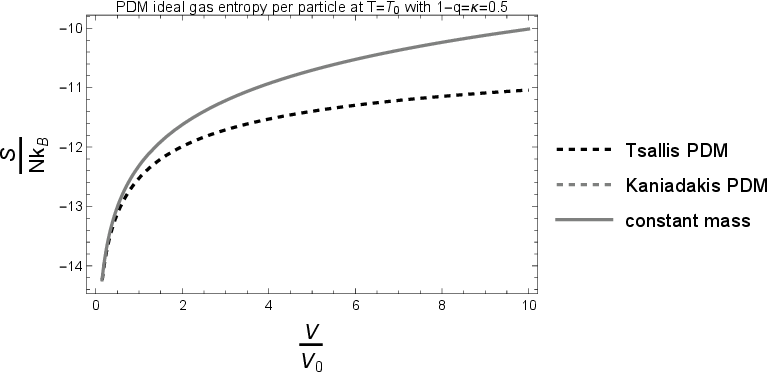}}
\vspace*{8pt}
\caption{Pressure (uper panel) and entropy (bottom panel) per particle in function of the volume at the temperature $T=T_0$ of the PDM ideal gas for the Tsallis, Kaniadakis and constant mass cases with $1-q=\kappa=0.5$.
\label{f1}}
\end{figure}

\begin{figure}[th]
\centerline{\includegraphics[width=9cm]{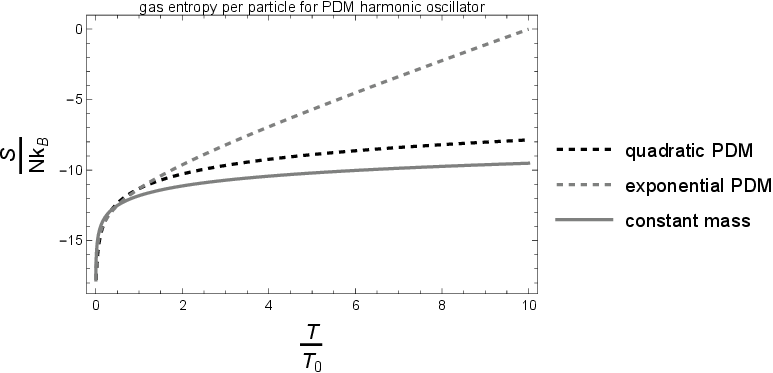}}
\vspace*{8pt}
\centerline{\includegraphics[width=9cm]{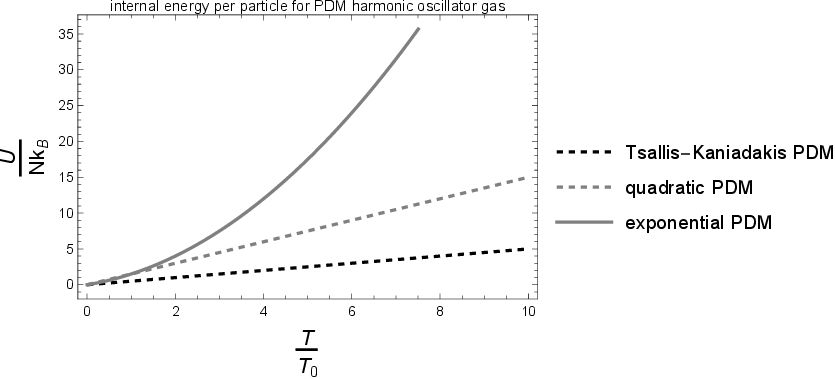}}
\vspace*{8pt}
\centerline{\includegraphics[width=9cm]{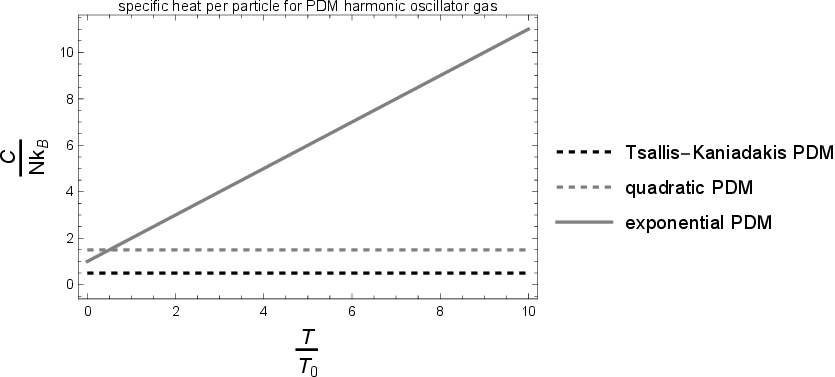}}
\vspace*{8pt}
\caption{Entropy (uper panel), internal energy (center panel) and specific heat (bottom panel) per particle in function of the temperature of a gas of non-interacting harmonic oscillators for the quadratic, exponential and constant mass cases with $c=0.5$.
\label{f2}}
\end{figure}

\subsection{Case 2: quadratic PDM $m(x)=m_0ax^2$ and harmonic potential $V(x)=\alpha x^2/2$}

By substituting $m(x)=m_0ax^2$ and $V(x)=\alpha x^2/2$ in \eqref{single-PDM-PF} and 
setting $(\sqrt{2\pi a m_0 (k_BT_0)^3/\alpha^2})/h=1$
(with $T_0$ a reference temperature),
we obtain the total partition function, by means of \eqref{PDM-PF}, as
\begin{eqnarray}
Z 
=\frac{1}{N!}\Bigg(\frac{T}{T_0}\Bigg)^{3N/2}. 
\end{eqnarray}
In this case, the thermodynamic functions (7a)--(7f) result in
\begin{subequations}
    \begin{align}
        A(V,T)=-k_BT \log \left(\frac{(T/T_0)^{\frac{3 N}{2}}}{N!}\right)\,,\\
        P(V,T)=0\,,\\
        S(V,T)=k_B\log \left(\frac{(T/T_0)^{\frac{3 N}{2}}}{N!}\right)+k_B\frac{3 N}{2}\,,\\
        G(V,T)=A(V,T)\,, \\
        U(V,T)=\frac{3 Nk_BT}{2}\,,
        \\        
        C(V,T)=\frac{3 Nk_B}{2}\,,
    \end{align}
\end{subequations}
which coincides with the corresponding ones to a three-dimensional gas of non-interacting harmonic oscillators of constant mass. 

\subsection{Case 3: exponential 
PDM $m(x)=m_0e^{-bx}$ and harmonic potential $V(x)=\alpha x^2/2$}

By substituting $m(x)=m_0e^{-bx}$ and $V(x)=\alpha x^2/2$ in \eqref{single-PDM-PF} and defining $c=b^2k_BT_0/(16\alpha)$ with the setting $\sqrt{2\pi^2m_0(k_BT_0)^2/\alpha}=1$, using \eqref{PDM-PF},
we obtain the total partition function
\begin{eqnarray}\label{case3-PF}
Z_{b,\alpha} 
=\frac{1}{N!}\Bigg(\frac{T}{T_0}e^{c(T/T_0)}\Bigg)^N\,.
\end{eqnarray}
In this case, the thermodynamic functions (7a)--(7f) result in 
\begin{subequations}
    \begin{align}
        A(V,T)=-k_B T \log \left(\frac{\left(\frac{T}{T_0} e^{c (T/T_0)}\right)^N}{N!}\right)\,,\\
        P(V,T)=0\,,\\
        S(V,T)=k_B\log \left(\frac{\left(\frac{T}{T_0} e^{c (T/T_0)}\right)^N}{N!}\right)+k_BN (c (T/T_0)+1)\,,\\ 
        G(V,T)=A(V,T)\,,\\
        U(V,T)=N k_B\frac{T}{T_0} \Bigg(c \frac{T}{T_0}+1\Bigg)\,,\\        C(V,T)=N k_B\Bigg(2c\frac{T}{T_0}+1\Bigg)\,.
    \end{align}
\end{subequations}
In Fig. 2, we show the entropy, internal energy and specific heat per particle
of a gas of non-interacting harmonic oscillators provided with a quadratic and exponential PDM for $c=0.5$. We see that the quadratic PDM harmonic oscillator gas exhibits the same behavior as the three-dimensional standard harmonic oscillator gas, thus showing a thermodynamical equivalence due to the freedom of choice of the pair $(m(x),V(x))$ in the context of the PDM canonical ensemble. The harmonic oscillator gas with a exponential PDM presents a linear increasing of the entropy for high temperatures, along with a quadratic behavior of the internal energy and a linear specific heat with the temperature. Thus, the employment of exponential effective masses could be useful to satisfy the Nernst's law of thermodynamics  $\lim_{T\rightarrow 0} C(V,T)=0$, 
by 
adjusting the value of $c$ in (23f).

\section{Conclusions}

We have presented a method for generating canonical partition functions of one-dimensional position-dependent systems, by means of a regime of fluctuations according to a delta function centered at a local inverse temperature dependent of the variable mass of the system.    
Our connection between superestatistical and PDM partition functions in the context of the canonical ensemble brings several new valuable aspects.

In Table 1, we summarize some relations between combinations of the investigated PDMs and potentials that illustrate the thermodynamic properties of PDM Tsallis and Kaniadakis ideal gases and of harmonic oscillator gases provided with quadratic and exponential PDMs. Thermodynamic equivalence between harmonic oscillator gas with quadratic PDM and the three dimensional ideal gas is also shown. 
The harmonic oscillator gas with exponential PDM presents a quadratic internal energy and linear specific heat with the temperature, that presumably could be used to satisfy the Nernst's law of thermodynamics in the context of PDM systems.  
Tsallis, Kaniadakis and standard ideal gases have the same internal energy and specific heat because the deformation only affects the position, i.e. the volume of the PDM ideal gas, and thus the dependence with the temperature remains unaltered. 
The inhomogeneities of the Tsallis and Kaniadakis PDMs imply an overall decreasing in the pressure of the ideal gas due to the fact that the particles acquire different inertia properties 
according to their positions. 
\begin{table}[]
    \centering
    \begin{tabular}{|c|c|c|c|c|}
        \hline
       PDM $m(x)$ & $V(x)$ & $\widetilde{V}(x)$ & $\gamma(x)$ & thermodynamics
        \\
        \hline
        $m_q(x)$ & $\equiv0$ & $\equiv0$ & $\equiv0$ & $P_q,P_\kappa<P_{\textrm{ideal gas}}$ \\ 
        & & & & $S_q=S_\kappa< S_{\textrm{ideal gas}}$\\
       $m_\kappa(x)$ & $\equiv0$ & $\equiv0$ & $\equiv0$ & 
        $U_q=U_\kappa= U_{\textrm{ideal gas}}$
        \\
        & & & & 
        $C_q=C_\kappa= C_{\textrm{ideal gas}}$
        \\
        \hline
        $m_0(1+ax)^2$ & $\alpha x^2/2$ & $\frac{\alpha x^2(1+ax)^2}{2}$ & $\frac{\beta_0}{(1+ax)^2}$ & $P=0$ \\
        & & & & $S=S_{\textrm{3D ideal gas}}$ \\
         & & & & $U=U_{\textrm{3D ideal gas}}$ \\
         & & & & $C=C_{\textrm{3D ideal gas}}$ \\
        \hline
        $m_0e^{-bx}$ & $\alpha x^2/2$ & $\frac{\alpha x^2e^{-bx}}{2}$ & $\beta_0e^{bx}$ & $P=0$\\
        & & & & $S\propto T$ for $T\gg T_0$ \\
        & & & & $U\propto T^2$ and $C\propto T$\\
        \hline
    \end{tabular}
    \caption{A table depicting the features of some PDM systems in the context of superestatistics along with their combinations of pairs of PDM masses-potentials and their thermodynamics.}
    \label{tab:my_label}
\end{table}
The interrelation between the constant mass, the PDM and the superestatistical partition functions in the canonical ensemble (p. f. stands for for partition functions), can be illustrated by
$$
\textrm{constant mass p. f.} \ \ \subseteq \ \
\textrm{PDM p. f.}
\ \ \subseteq \ \
\textrm{superestatistical p. f.}
$$
Finally, 
we conjecture that combined 
PDM and superestatistical scenarios could shed light to 
study more deeply 
the thermodynamics of heteregeneous structures.

\section*{Acknowledgments}
Ignacio S. Gomez and Ronaldo Thibes acknowledge support from the Department of Exact
and Natural Sciences of the State University of Southwest
Bahia (UESB), Itapetinga, Bahia, Brazil and from the PROFÍSICA (UESC), Ilhéus, Bahia, Brazil. 
Ignacio S. Gomez acknowledges support received from the Conselho Nacional de Desenvolvimento Científico e Tecnológico (CNPq), Grant Number
316131/2023-7.
Daniela de Almeida dos Santos acknowledges support received as scientific initiation scholarship fellow of the UESB.
Matheus Gabriel Alves Santos acknowledges support received as scientific initiation scholarship fellow of the CNPq.

\end{document}